# Quantum Discord Dynamics for Two-Level Atom Initially in Thermal Equilibrium Interacting with n-Photon State

Hari Prakash[1,2] and Manoj K Mishra[1]

[1]Physics Department, University of Allahabad, Allahabad, India

[2]Indian Institute of Information Technology, Allahabad, India

e-mail: prakash_hari123@rediffmail.com, hariprakash@iiita.ac.in, manoj.qit@gmail.com

**Abstract**

We investigate the quantum discord dynamics and inversion operator of a two-level atom initially in thermal equilibrium mixed state interacting with a cavity field prepared in a n-photon Fock state. We considered the Jaynes-Cummings model which is an exactly quantum mechanical model of interaction between two different quantum systems, an optical cavity in single mode n-photon Fock state and a two-level atom. The influence of interaction time and measurement basis used for calculating the conditional entropy is discussed on the evolution of the quantum discord and on its minimum. The evolution of mean value of inversion operator with interaction is also discussed. It is found that the quantum discord and inversion both shows oscillatory behaviour and also shows the phenomenon of beats with interaction time. For a given initial cavity photon Fock state the beat period of minimum quantum discord is one half of that for inversion operator. However, as the number of photons in cavity increases the frequency of oscillations of quantum discord and inversion increases.

## 1 Introduction

The interaction between atoms and the electromagnetic field is one of the most interesting problem which has been extensively studied in near past. The Jaynes-Cumming model (JCM) [1] is an exactly soluble and simplest quantum mechanical model of interaction between two different quantum systems- a single two-level atom and a single mode quantized electromagnetic field. It is simple to solve, hence allows the study of quantum mechanical properties associated with atom and field analytically. The JCM model can be experimentally realized [2].

The atomic systems have found new application in quantum information processing [3-5]. The main requirement in quantum information processing task is the quantum entanglement [6]. The interaction of atoms with cavity field has been shown to be an efficient source of atom-atom, atom-cavity, and cavity-cavity entanglement [7-11]. Appreciable number of studies related to the study of influence of atom-field interaction on the entanglement between atom and field has been done. The reason behind the study of quantum entanglement in such system is due to its fundamental nature and its applicability. However as discussed in previous chapter quantum entanglement do not exhausts the complete quantum correlations. Zurek et al [12, 13] proposed quantum discord as a measure of quantum correlation defined as, discrepancy between two expressions of mutual information, one obtained by using conditional entropy and the other by performing local measurement on any one of the subsystem, classically these two mutual information are equivalent. Since QD is based on the total correlation (mutual information), it is capable to sense correlation in non-separable (entangled) as well as in separable states. Appreciable amount of works related to theoretical development of QD [14-30], its dynamical property under the effect of decoherence [31-37] and its comparison with QE have been done over the past decade [38, 39]. Quantum discord acts as main resource in many quantum information processing tasks like quantum locking of classical correlations [40], quantum computation with out entanglement [41, 42] and quantum teleportation without entanglement [43].

In present chapter we consider a model consisting of a single two-level atom initially in thermal equilibrium mixed state interacting with a cavity field initially prepared in n-photon Fock state and investigate the dynamics of quantum discord and inversion operator.

## 2 Model of the atom-cavity system

The Hamiltonian of the JCM which describes a system of two-level atom, consisting of the states $|0\rangle$ and $|1\rangle$, coupled to a single mode exactly resonant radiation field cavity, in the rotating wave approximation is given by

$$H_{ac} = \omega a^\dagger a + \frac{\omega_0}{2}\sigma_z + \beta(a^\dagger \sigma_- + a\sigma_+) \tag{1}$$

where suffix *ac* refers to atom-cavity interaction, $\omega$ is the common frequency of the atom and the cavity, $a^\dagger(a)$ is the creation (annihilation) operator of the field mode, $\beta$ is the atom field coupling constant, $\sigma_\pm$ and $\sigma_z$ are the atomic raising, lowering, and inversion operators of the atom.

$$\sigma_+ = \begin{pmatrix} 0 & 1 \\ 0 & 0 \end{pmatrix}, \quad \sigma_- = \begin{pmatrix} 0 & 0 \\ 1 & 0 \end{pmatrix}, \quad \sigma_z = \begin{pmatrix} 1 & 0 \\ 0 & -1 \end{pmatrix}. \tag{2}$$

The first and second terms in equation (1) represent the free field and free atom Hamiltonian while third term represents the interaction Hamiltonian $H_{ac}^{int}$

$$H_{ac}^{int} = \beta(a^\dagger R_- + a R_+).$$

The time evolution operator in the interaction picture is described by

$$U(t) = \exp(-i H_{ac}^{int} t). \tag{3}$$

The effect of time evolution on a system of two-level atom interacting with n-photon cavity state initially in states, $|0,n\rangle$ and $|1,n\rangle$, is described by

$$\Psi(t) = U(t)|0,n\rangle = C_n|0,n\rangle - i S_n|1,n-1\rangle \tag{4}$$

$$\Psi(t) = U(t)|1,n\rangle = C_n|1,n\rangle - i S_n|0,n+1\rangle \tag{5}$$

where

$$\begin{aligned} C_n &= \cos\psi_n, \; S_n = \sin\psi_n, \\ C_{n+1} &= \cos\psi_{n+1}, \; S_{n+1} = \sin\psi_{n+1}, \\ \psi_n &= \beta t \sqrt{n}, \; \psi_{n+1} = \beta t \sqrt{n+1} \end{aligned} \tag{6}$$

We assume that initially atom is in thermal equilibrium represented by density matrix

$$\rho_a(0) = \lambda_0 |0\rangle\langle 0| + \lambda_1 |1\rangle\langle 1|, \tag{7}$$

where,

$$\lambda_0 = [1 + e^{-\omega/KT}], \quad \lambda_1 = e^{-\omega/KT}/[1 + e^{-\omega/KT}] \tag{8}$$

and the initial state of cavity is n-photon state given by density matrix

$$\rho_c(0) = |n\rangle\langle n|. \tag{9}$$

Initially atom-cavity has no quantum correlation, i.e., the atom-cavity joint state is a product state and write

$$\rho_{ac}(0) = \rho_a(0) \otimes \rho_c(0) \tag{10}$$

Using equation (4) and (5), the state of the atom-cavity system at any arbitrary time is described by

$$\begin{aligned} \rho_{ac}(t) = U(t)\rho_{ac}(t)U^\dagger(t) &= \lambda_0(C_n^2|0,n\rangle\langle 0,n| + S_n^2|1,n-1\rangle\langle 1,n-1|) \\ &+ \lambda_1(C_{n+1}^2|1,n\rangle\langle 1,n| + S_{n+1}^2|0,n+1\rangle\langle 0,n+1|) \\ &+ i\lambda_0 C_n S_n (|0,n\rangle\langle 1,n-1| - |1,n-1\rangle\langle 0,n|) \\ &+ i\lambda_1 C_{n+1} S_{n+1}(|1,n\rangle\langle 0,n+1| - |0,n+1\rangle\langle 1,n|) \end{aligned} \tag{11}$$

In density matrix form this can be written as

$$\rho_{ac}(t) = U(t)\rho_{ac}(t)U^{\dagger}(t)$$

$$= \begin{bmatrix} 0 & 0 & 0 & 0 & 0 & 0 \\ 0 & \lambda_0 C_n^2 & 0 & i\lambda_0 C_n S_n & 0 & 0 \\ 0 & 0 & \lambda_1 S_{n+1}^2 & 0 & -i\lambda_1 C_{n+1} S_{n+1} & 0 \\ 0 & -i\lambda_0 C_n S_n & 0 & \lambda_0 S_n^2 & 0 & 0 \\ 0 & 0 & i\lambda_1 C_{n+1} S_{n+1} & 0 & \lambda_1 C_{n+1}^2 & 0 \\ 0 & 0 & 0 & 0 & 0 & 0 \end{bmatrix} \quad (12)$$

The eigenvalues of density matrix $\rho_{ac}(t)$ are given by

$$\rho_{ac}(t) \equiv \{\lambda_0, \lambda_1, 0, 0, 0, 0\} \quad (13)$$

The reduced density matrix of atom is given by

$$\rho_a(t) = \begin{bmatrix} \lambda_0 C_n^2 + \lambda_1 S_{n+1}^2 & 0 \\ 0 & \lambda_0 S_n^2 + \lambda_1 C_{n+1}^2 \end{bmatrix}. \quad (14)$$

The eigenvalues of density matrix $\rho_a(t)$ are given by,

$$\rho_a(t) \equiv \{\lambda_0 C_n^2 + \lambda_1 S_{n+1}^2, \lambda_0 S_n^2 + \lambda_1 C_{n+1}^2\} \quad (15)$$

It is to be noted here that at an arbitrary time the cavity we choose can have finite dimensionality more than two. In our case it is of three dimension represented by orthogonal photonic state, $|n-1\rangle$, $|n\rangle$, and $|n+1\rangle$; and we shall see that this will give us some fascinating results about the dynamics of quantum discord.

## 3 Quantum discord

To study the dynamics of quantum discord we perform an ideal von Neumann projection measurement on atom by a complete set of one-dimensional projector

$$|\pi_0^a\rangle = C|0\rangle + zS|1\rangle, |\pi_1^a\rangle = z^*S|0\rangle - C|1\rangle, \quad (16)$$

where

$$C = \cos\theta, \; S = \sin\theta, \; z = \exp(i\phi) \quad (17)$$

and satisfies the completeness relation, $\sum_{j=0,1}|\pi_j^a\rangle\langle\pi_j^a| = I$. Cavity state after measurement on atom corresponding to outcomes $\{\pi_j^a\}$ is

$$\rho_{c|\pi_j^a}(t) = [Tr_a(|\pi_j^a\rangle\langle\pi_j^a|\rho_{ac}(t)|\pi_j^a\rangle\langle\pi_j^a|)]/P_j, \quad (18)$$

where $P_j$ is the probability of outcome $\pi_j^a$ given by

$$P_j = Tr[|\pi_j^a\rangle\langle\pi_j^a|\rho_{ac}(t)] \quad (19)$$

Using this we get

$$\rho_{c|\pi_0^a}(t) = \frac{1}{p_0}\begin{bmatrix} \lambda_0 S^2 S_n^2 & -iz^*\lambda_0 CSC_n S_n & 0 \\ iz\lambda_0 CSC_n S_n & \lambda_0 C^2 C_n^2 + \lambda_1 S^2 C_{n+1}^2 & iz^*\lambda_1 CSC_{n+1} S_{n+1} \\ 0 & -iz\lambda_1 CSC_{n+1} S_{n+1} & \lambda_1 C^2 S_{n+1}^2 \end{bmatrix} \quad (20)$$

$$\rho_{c|\pi_1^a}(t) = \frac{1}{p_1}\begin{bmatrix} \lambda_0 C^2 S_n^2 & iz^*\lambda_0 CSC_n S_n & 0 \\ -iz\lambda_0 CSC_n S_n & \lambda_0 S^2 C_n^2 + \lambda_1 C^2 C_{n+1}^2 & -iz^*\lambda_1 CSC_{n+1} S_{n+1} \\ 0 & iz\lambda_1 CSC_{n+1} S_{n+1} & \lambda_1 S^2 S_{n+1}^2 \end{bmatrix} \quad (21)$$

where

$$P_0 = \lambda_0(C^2 C_n^2 + S^2 S_n^2) + \lambda_1(C^2 S_{n+1}^2 + S^2 C_{n+1}^2) \quad (22)$$

$$P_1 = \lambda_0(S^2 C_n^2 + C^2 S_n^2) + \lambda_1(S^2 S_{n+1}^2 + C^2 C_{n+1}^2) \quad (23)$$

Eigenvalues of the states $\rho_{c|\pi_0^a}(t)$ and $\rho_{c|\pi_1^a}(t)$, are given by

$$\rho_{c|\pi_0^a}(t) \equiv \{0, \tfrac{1}{2}(1\pm\sqrt{1-4y^{\pi_0}})\} \quad (24)$$

$$\rho_{c|\pi_1^a}(t) \equiv \{0, \tfrac{1}{2}(1\pm\sqrt{1-4y^{\pi_1}})\} \quad (25)$$

where

$$y^{\pi_0} = [\lambda_0\lambda_1(C^4 C_n^2 S_{n+1}^2 + S^4 S_n^2 C_{n+1}^2 + C^2 S^2 S_n^2 S_{n+1}^2)]/P_1^2 \quad (26)$$

$$y^{\pi_1} = [\lambda_0\lambda_1(S^4 C_n^2 S_{n+1}^2 + C^4 S_n^2 C_{n+1}^2 + C^2 S^2 S_n^2 S_{n+1}^2)]/P_2^2 \quad (27)$$

It is to be noted that none of the eigenvalues of $\rho_{c|\pi_0^a}(t)$ and $\rho_{c|\pi_1^a}(t)$, depends upon the phase angle $\phi$ of the measurement basis, thus whatever be the value of quantum discord it will be independent of the phase angle $\phi$ of the measurement basis. Quantum discord is defined as difference between the mutual informations $I(c:a)$ and $J(c:a)_{\{\pi_j^a\}}$, based on joint entropy and conditional entropy, respectively, as discussed in chapters 6 and 1, and can be written as

$$D(c:a)_{\{\Pi_j^a\}} = I(c:a) - J(c:a)_{\{\pi_j^a\}} = S(\rho_a) - S(\rho_{ac}) + S(\rho_{c|\{\Pi_j^a\}}), \tag{28}$$

where

$$S(\rho) = -Tr(\rho \log \rho), \tag{29}$$

is the von Neumann entropy for a state represented by density matrix $\rho$ and conditional entropy is given by

$$S(\rho_{\{c|\pi_j^a\}}(t)) \equiv \sum_j P_j S(\rho_{c|\pi_j^a}). \tag{30}$$

We are also interested in the minimum value of discord which can be obtained by minimizing its expression (28) over all possible ideal von Neumann projection measurement and is given by relation

$$\delta(X:Y)_{\{\Pi_j^Y\}} = \min_{\{\Pi_j^Y\}} [D(X:Y)_{\{\Pi_j^Y\}}]. \tag{31}$$

The entropies $S(\rho_a)$ and $S(\rho_{ac})$ can be obtained by using the eigenvalues of the density matrices $\rho_a$ and $\rho_{ac}$, given in equations (15) and (13), respectively, in equation (29). The conditional entropy $S(\rho_{\{c|\pi_j^a\}}(t))$ can be obtained by using its eigenvalues given in equation (24) and (25) in equation (30). Using these entropies in equation (28), the expression for quantum discord is given by

$$\begin{aligned}
D(c:a) = &-\{(\lambda_0 C_n^2 + \lambda_1 S_{n+1}^2)\log(\lambda_0 C_n^2 + \lambda_1 S_{n+1}^2) + (\lambda_0 S_n^2 + \lambda_1 C_{n+1}^2) \\
&\times \log(\lambda_0 S_n^2 + \lambda_1 C_{n+1}^2)\} + \{\lambda_0 \log \lambda_0 + \lambda_1 \log \lambda_1\} \\
&- \left[\frac{P_0}{2}(1+\sqrt{1-4y^{\pi_0}})\log(1+\sqrt{1-4y^{\pi_0}}) + (1-\sqrt{1-4y^{\pi_0}})\log(1-\sqrt{1-4y^{\pi_0}})\} \right. \\
&\left. + \frac{P_1}{2}\{(1+\sqrt{1-4y^{\pi_1}})\log(1+\sqrt{1-4y^{\pi_1}}) + (1-\sqrt{1-4y^{\pi_1}})\log(1-\sqrt{1-4y^{\pi_1}})\}\right]
\end{aligned} \tag{32}$$

where $P_j$ and $y^{\pi_j}$ are given in equations (22, 23, 26 and 27).

We can write a Matlab code to calculate the quantum discord as mentioned in equation (32) and to minimize the calculated value of discord over all possible ideal von Neumann projection measurement (i.e., minimization by running over all values of measurement parameter $\theta$). Fig 1 shows the variation quantum discord with respect to interaction time and measurement basis parameter $\theta$ for different initial photonic Fock state of cavity, while Fig. 2 shows the variation of minimum of quantum discord with respect to interaction time for different initial photonic Fock state of cavity. Another important quantity is the population inversion ($\langle \sigma_z \rangle$) defined as the mean value of inversion operator ($\sigma_z$), described by relation

$$\langle \sigma_z \rangle = Tr[\rho_{ac}(t)\sigma_z]. \tag{33}$$

Using equation (11) in equation (33), we get,

$$\langle \sigma_z \rangle = \lambda_0(C_n^2 - S_n^2) - \lambda_1(C_{n+1}^2 - S_{n+1}^2) \tag{34}$$

We plotted the mean value of inversion operator in Fig. 2, with respect to interaction time for different initial photonic Fock state of cavity.

## 4 Results and discussion

Consider the case $\lambda_0 = \lambda_1 = 0.5$, i.e., at initial time ($t=0$) the state of the atom is $\rho_a(0) = (1/2)(|0\rangle\langle 0| + |1\rangle\langle 1|)$, that correspond to the limit of very high temperatures. With this initial condition on atom, Figs 1(a), (b), (c), and (d) show the contour plot of quantum discord with respect to interaction time ($\beta t$) and measurement parameter $\theta$, for initial photonic Fock state of cavity with n =1, 2, 4 and 8 numbers of photons, respectively. Fig. 1 (a) shows the interesting case that there are certain ranges of interaction times for which the quantum correlation or discord between atom and the cavity do not vanish for any measurement basis (i.e., for any value of measurement parameter $\theta$). Figs. 1 (a-d) also show that, for any measurement basis, the quantum discord is rapidly oscillating in nature in a complex way with respect to interaction time. However, the frequency of oscillation increases with increase in numbers of photon initially in cavity. This is also evident by the fact that the density of the fringes that represents the variation of quantum discord increases as the number of photons increase from n=1 to n=8.

We also note that the quantum discord is showing phenomenon of beats with interaction time. Figs. 2 (a-e) show the minimum value of quantum discord over the measurement basis and the mean value of inversion operator for n=1, 2, 4, 8, 15 with respect to interaction time. It is clear that both the minimum quantum discord and inversion operator exhibit the phenomenon of beats with interaction time. The frequency of oscillation increases rapidly with increasing number of photons.

The beating of $\langle \sigma_z \rangle$ can be explained easily by equation (34), which simplifies for the case $\lambda_0 = \lambda_1 = 1/2$ to

$$\langle \sigma_z \rangle = \sin(\beta t[\sqrt{n+1} + \sqrt{n}])\sin(\beta t[\sqrt{n+1} - \sqrt{n}]).$$

The first factor involves larger frequency and gives phase of the oscillations. The mean frequency of these oscillations is $\beta(\sqrt{n+1} + \sqrt{n})$ and the mean periodic time is $2\pi/\beta(\sqrt{n+1} + \sqrt{n})$. The amplitude of oscillations oscillates itself periodically and the beat frequency is $2\beta(\sqrt{n+1}-\sqrt{n}) = 2\beta/(\sqrt{n+1}+\sqrt{n})$ and the beat period is $(\pi/\beta)(\sqrt{n+1}+\sqrt{n})$. In one beat-period number of oscillations is $(\sqrt{n+1}+\sqrt{n})/(\sqrt{n+1}-\sqrt{n}) = \frac{1}{2}[2n+1+2\sqrt{n(n+1)}]$. For $n \gg 1$, the periodic time and the beat period are $(\pi/\beta)\sqrt{n}$ and $(2\pi/\beta)\sqrt{n}$, respectively, i.e., during two minima of amplitude of the oscillations, about 2n oscillations take place. The cases for n=1, 2, 4, 8 and 15 are shown in Figs 2(a-d). For largest $n$, $n = 15$ in Fig 2d, the number of oscillations in one beat period is nearly 31 while the expression $\frac{1}{2}[2n+1+2\sqrt{n(n+1)}]$ gives 30.99 and the approximate result gives $2n = 30$ oscillations.

Quantum discord for the problem in hand is a really very complicated function of $n$, $\theta$ and the interaction time $\beta t$ but still permits some general observations. Quantum discord $D$ is a periodic function of $\theta$ with the period $\pi/2$. It admits maxima or minima only at $\theta = 0, \pi/4, \pi/2, 3\pi/4$ and $\pi$. There is no general periodicity with the interaction time. The occurrence of maximum (minimum) at $0, \pi/2$ or $\pi$ ($\pi/4$ or $3\pi/4$) at a given $\beta t$ changes suddenly to $\pi/4$ or $3\pi/4$ ($0, \pi/2$ or $\pi$) on changing $\beta t$ without any simple prescription. This has been studied explicitly for $n = 8$ and is shown in Fig. 2d.

The quantum Discord $D$ plotted against interaction time $\beta t$ shows oscillatory behaviour with amplitude of oscillations also showing oscillations. The periodic time of oscillations and beat period shows interesting variation with $\theta$. For $\theta = \pi/2$ and $n = 8$ behaviour of $D$ is shown in Fig. 3. For $\theta = \pi/2$ oscillations of $D$ occur with a period which is one half of the period for $\langle\sigma_z\rangle$. The beat period is also one half of the corresponding value for $\langle\sigma_z\rangle$. For $\theta = \pi/4$ the oscillatory behaviour of $D$ is very complicated and interesting. For some intervals the period is same as that for $\langle\sigma_z\rangle$ while for remaining intervals it is one half of that for $\langle\sigma_z\rangle$.

For minimum (against $\theta$) quantum discord $\delta$, it is seen that its behaviour is oscillatory with $\beta t$ and there appear beats in the amplitude. The period of oscillation of $\delta$ is about one half of that for oscillations of $\langle\sigma_z\rangle$ as is clear from Figs &.2 (a)-(e). The beat period of $\delta$ is also about one half of that for $\langle\sigma_z\rangle$. One other interesting observation is that, for even beats, the maxima of $\delta$ are alternately high and low although this behaviour is absent for minima (see Fig. 2(d) and (e) for $n = 8$ or 15). In addition to this we also see that the high maxima in even beats show a dip.

It remains that the results involve complicated functions and it is not easy to correlate the general observations with the formula.

**Acknowledgement:**

We are grateful to Prof. N. Chandra and Prof. R. Prakash for their interest and stimulating discussions. Discussions with Ajay Kumar Yadav, Ajay Kumar Maurya and Vikram Verma are gratefully acknowledged. One of the authors (MKM) acknowledges the UGC for financial support under UGC-SRF fellowship scheme.

**Figure 1**

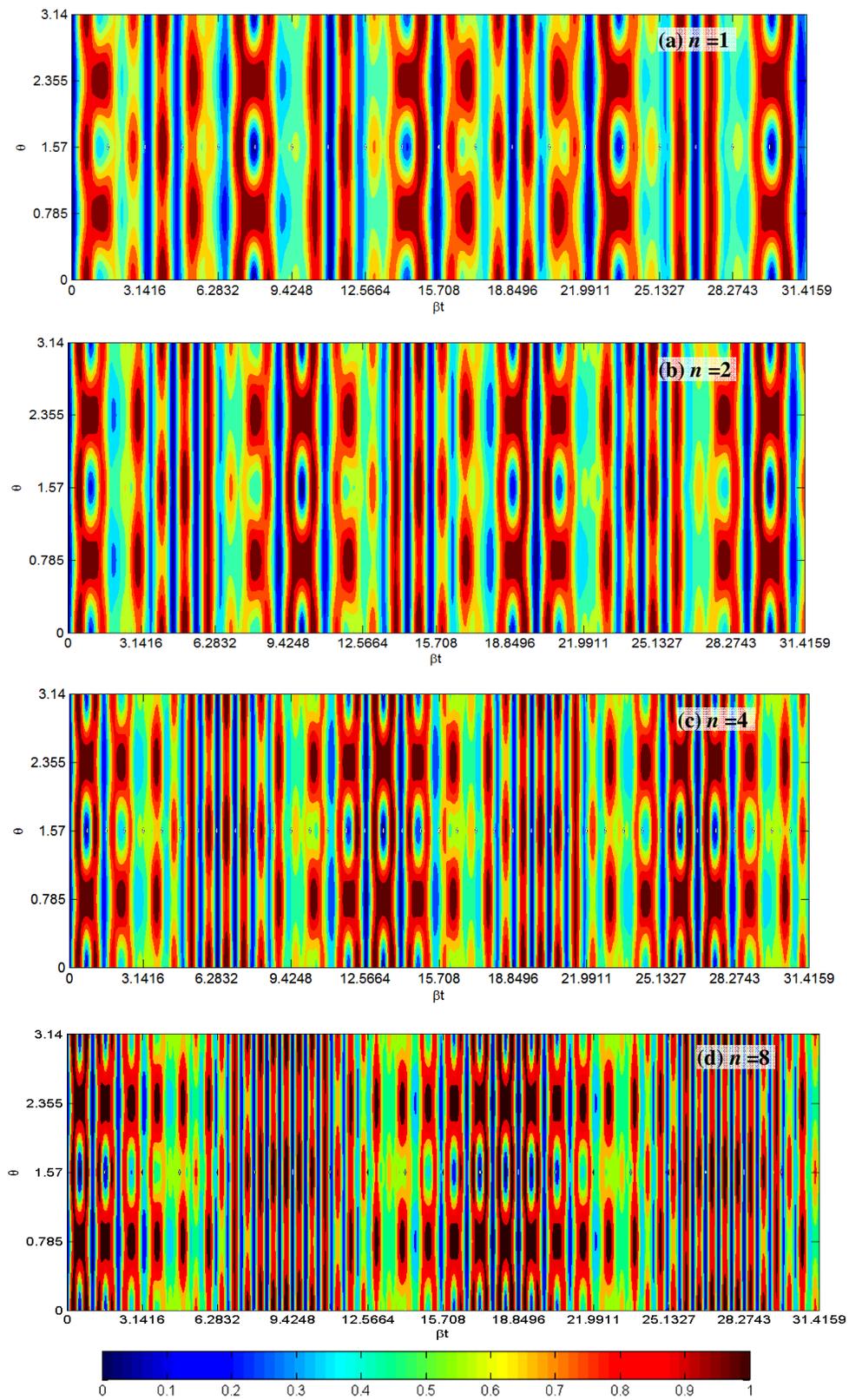

Fig.1 Quantum discord D(c:a) with respect to interaction time and measurement parameter for n=1,2,4,8 number of photons initially in cavity.

**Figure 2**

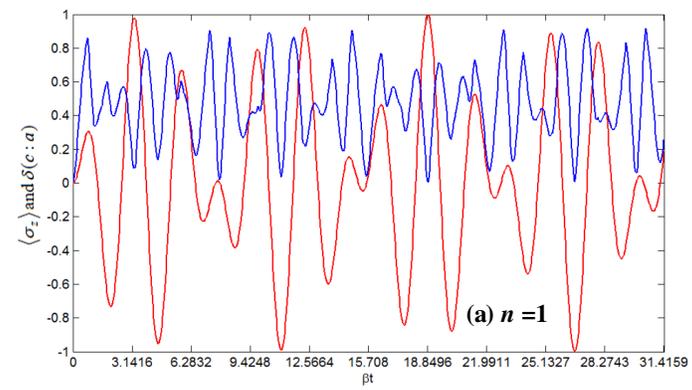

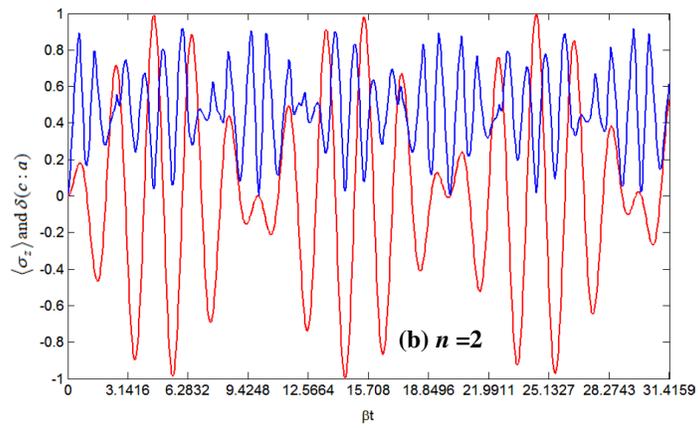

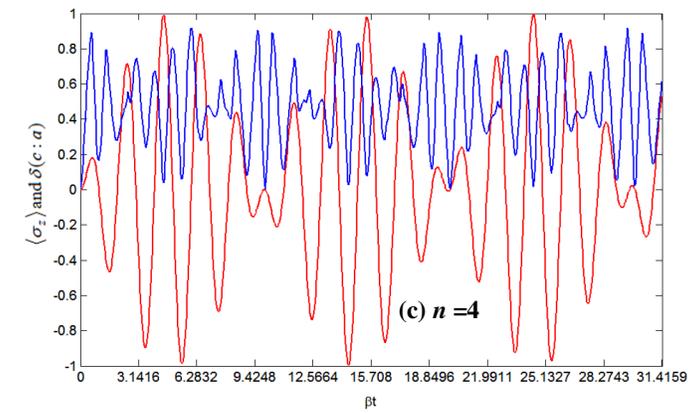

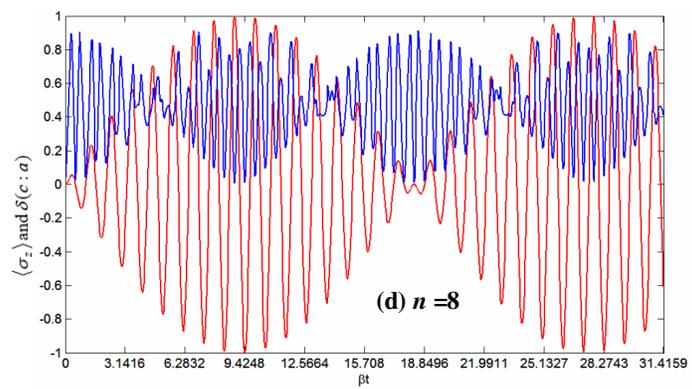

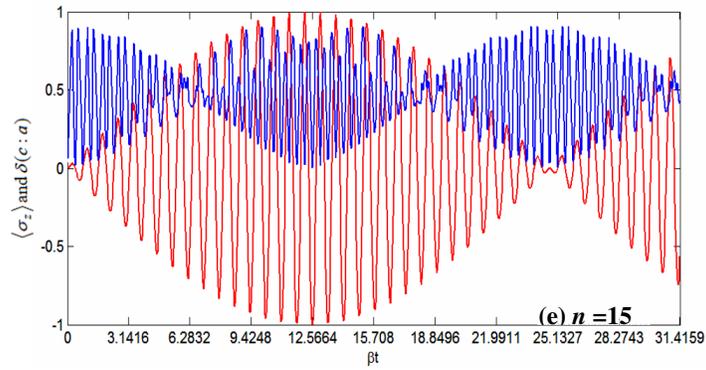

Fig 2 Minimum value of Quantum discord δ(c:a) (blue) and mean value of inversion operator (red) with respect to interaction time for $n$ =1, 2, 4, 8, 15 number of photons initially in cavity.

**Figure 3**

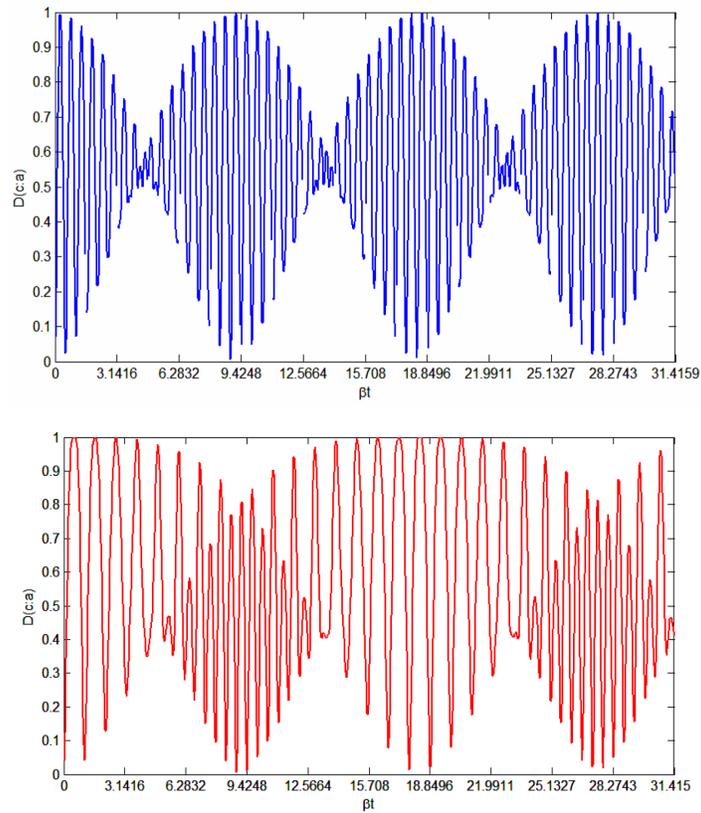

Fig 3 Quantum discord δ(c:a) for $\theta = \pi/2$ (blue) and $\theta = \pi/4$ (red) with respect to interaction time for n=8 number of photons initially in cavity.